\documentclass[twocolumn,showpacs,pra,aps,superscriptaddress]{revtex4}
\usepackage{array}
\usepackage{amsmath}
\usepackage{graphicx}
\usepackage{dsfont}
\usepackage{color}
\def\bra#1{\langle #1|}
\def\ket#1{|#1 \rangle}

\begin{document}

\title{Witnessing entanglement in phase space using inefficient detectors}

\author{Seung-Woo Lee}
\affiliation{Clarendon Laboratory,
University of Oxford, Parks Road, Oxford OX1 3PU, United Kingdom}

\author{Hyunseok Jeong}

\affiliation{Center for Subwavelength Optics and Department of
Physics and Astronomy, Seoul National University, Seoul, 151-742,
Korea}

\author{Dieter Jaksch}

\affiliation{Clarendon Laboratory, University of Oxford, Parks Road,
Oxford OX1 3PU, United Kingdom} \affiliation{Center for Quantum
Technologies, National University of Singapore, Singapore 117543,
Singapore}

\date{\today}

\begin{abstract}
We propose a scheme for witnessing entanglement in phase space by
significantly inefficient detectors. The implementation of this
scheme does not require any additional process for correcting errors
in contrast to previous proposals. Moreover, it allows us to detect
entanglement without full {\em a priori} knowledge of the detection
efficiency. It is shown that entanglement in single photon entangled
and two-mode squeezed states can be witnessed with detection
efficiency as low as 40\%. Our approach enhances the possibility of
witnessing entanglement in various physical systems using current
detection technologies.
\end{abstract}


\maketitle

\section{Introduction}

Entanglement is one of the most remarkable features of quantum
mechanics which can not be understood in the context of classical
physics. It has been shown that entanglement can exist in various
physical systems and play a role in quantum phenomena
\cite{Amico-RMP-2008,Vedral-Nature-2008}. Moreover, its properties
can be used as a resource for quantum information technologies such
as quantum computing, quantum cryptography, and quantum
communication \cite{Nielsen-Cam-2000}. Therefore, detecting
entanglement is one of the most essential tasks both for studying
fundamental quantum properties and for applications in quantum
information processing. Although various entanglement detection
schemes have been proposed \cite{Horodecki-RMP-2009}, their
experimental realization suffers from imperfections of realistic
detectors since measurement errors wash out quantum correlations.
This difficulty becomes more significant with increasing system
dimensionality and particularly in continuous variable systems where
entanglement is increasingly attracting interest
\cite{Braunstein-RMP-2005}.

Quantum tomography provides a method to reconstruct complete
information of quantum states in phase space formalism
\cite{Vogel-PRA-1989,Banaszek-PRL-1996,Lvovsky-RMP-2009}. The
reconstructed data can be used to determine whether the state is
entangled or not with the help of an entanglement witness (EW). Bell
inequalities that were originally derived for discriminating quantum
mechanics from local realism \cite{Bell-Physics-1964} can also be
used for witnessing entanglement since their violation guarantees
the existence of entanglement. Banaszek and W\'{o}dkiewicz (BW)
\cite{Banaszek-PRA-1998} suggested a Bell-type inequality (referred
to BW-inequality in this paper) which can be tested by way of
reconstructing the Wigner function at a few specific points of phase
space. However, imperfections of tomographic measurements constitute
a crucial obstacle for its practical applications. Several schemes
have been considered to overcome this problem
\cite{Lvovsky-RMP-2009} such as numerical inversion
\cite{Kiss-PRA-1995} and maximum-likelihood estimation
\cite{Banaszekb-PRA-1998}, but they require a great amount of
calculations or iteration steps for high dimensional and continuous
variable systems.

In this paper we propose an alternative entanglement detection
scheme in phase space formalism, which can be used in the presence
of detection noise. We formulate an entanglement witness (EW) in the
form of a Bell-like inequality using the experimentally measured
Wigner function. For this, we include the effects of detection
efficiency into possible measurement outcomes. Possible expectation
values of the entanglement witness are bounded by the maximal
expectation value when separable states are assumed. Any larger
expectation value guarantees the existence of entanglement.

Our approach shows the following remarkable features: (i) in
contrast to previous proposals
\cite{Lvovsky-RMP-2009,Banaszekb-PRA-1998} it does not require any
additional process for correcting measurement errors; (ii) it allows
us to witness entanglement e.g. in single-photon entangled and
two-mode squeezed states with detection efficiency as low as 40\%;
(iii) our scheme is also valid when the precise detection efficiency
is not known prior to the test; (iv) finally, we note that our
approach is applicable to detect any quantum state represented in
phase space formalism.

\section{Observable associated with efficiency}

We begin by introducing an observable associated with the detection
efficiency $\eta$ and an arbitrary complex variable $\alpha$:
\begin{eqnarray}
 \label{eq:ObOp}
  \hat{O}(\alpha) =
  \begin{cases}\frac{1}{\eta}
\hat{\Pi}(\alpha)+(1-\frac{1}{\eta})\openone & \text{if $\frac{1}{2} < \eta \leq 1$},\\
2\hat{\Pi}(\alpha)-\openone & \text{if $ \eta \leq \frac{1}{2}$},
\end{cases}
\end{eqnarray}
where $\hat{\Pi}(\alpha) =
\sum^{\infty}_{n=0}(-1)^{n}\ket{\alpha,n}\bra{\alpha,n}$ is the
displaced parity operator and $\openone$ is the identity operator.
$\ket{\alpha,n}=\hat{D}(\alpha)\ket{n}$ is the displaced number
state produced by applying the Glauber displacement operator $\hat
D(\alpha)$ to the number state $|n\rangle$.

Let us then consider the expectation value of observable
(\ref{eq:ObOp}) when the measurement is carried out with efficiency
$\eta$. In general, measurement errors occur when not all particles
are counted in the detector. Thus the real probability distribution
of particles, $P(n)$, transforms to another distribution,
$P_{\eta}(m)$, by the generalized Bernoulli transformation
\cite{Leonhardt-1997}:
$P_{\eta}(m)=\sum^{\infty}_{n=m}P(n)\binom{n}{m}(1-\eta)^{n-m}\eta^m$.
Thus the expectation value of the parity operator is obtained as
\begin{eqnarray}
\label{eq:parityexp} \langle
\hat{\Pi}(\alpha)\rangle_{\eta}=\sum^{\infty}_{m=0}(-1)^mP_{\eta}(\alpha,m)
=\sum^{\infty}_{n=0}(1-2\eta)^n P(\alpha,n),
\end{eqnarray}
where $\langle \cdot \rangle_{\eta}$ implies the statistical average
measured with efficiency $\eta$. Here $P_{\eta}(\alpha,m)$ and
$P(\alpha,n)$ are the measured and real particle number
distributions in the phase space displace by $\alpha$, respectively.

We define the Wigner function experimentally measured with
efficiency $\eta$ as
\begin{eqnarray}
\label{eq:Wigexp} W^{\eta}(\alpha)&\equiv& \frac{2}{\pi}\langle
\hat{\Pi}(\alpha)\rangle_{\eta},
\end{eqnarray}
which is given as a rescaled Wigner function by Gaussian smoothing.
Note that a smoothed Wigner function can be identified with a
$s$-parameterized quasi-probability function as
$W^{\eta}(\alpha)=W(\alpha;-(1-\eta)/\eta)/\eta$ where $W(\alpha;
s)= (2/(\pi(1-s)))\sum^{\infty}_{n=0}((s+1)/(s+1))^n P(\alpha,n)$
\cite{Cahill-PR-1969}. This identification is available both for
homodyne \cite{Vogel-PRA-1989,Leonhardt-PRA-1993} and number
counting tomography methods \cite{Banaszek-PRL-1996}. After series
of measurements with efficiency $\eta$, we can obtain the
expectation value of the observable (\ref{eq:ObOp}) as
\begin{eqnarray}
 \label{eq:expObOp}
  \langle \hat{O}(\alpha) \rangle_{\eta}=
  \begin{cases}\frac{\pi}{2\eta}
W^{\eta}(\alpha)+1-\frac{1}{\eta} & \text{if $\frac{1}{2} < \eta \leq 1$},\\
\pi W^{\eta}(\alpha)-1 & \text{if $ \eta \leq \frac{1}{2}$},
\end{cases}
\end{eqnarray}
which is bounded as $|\langle \hat{O}(\alpha) \rangle_{\eta}|\leq 1$
for all $\eta$.

\section{Entanglement witness in phase space}

Let us formulate an entanglement witness (EW) in the framework of
phase space. Suppose that two separated parties, Alice and Bob,
measure one of two observables, denoted by $\hat{A}_1$, $\hat{A}_2$
for Alice and $\hat{B}_1$, $\hat{B}_2$ for Bob. All observables are
variations of the operator (\ref{eq:ObOp}) as $\hat{A}_a
=\hat{O}(\alpha_a)$ and $\hat{B}_a =\hat{O}(\beta_b)$ with
$a,b=1,2$. We then formulate a Hermitian operator as a combination
of each local observable $\hat{A}_a$, $\hat{B}_b$ in the form
\begin{eqnarray}
 \label{eq:BOp}
  {\cal \hat{W}}= \hat{C}_{1,1}+\hat{C}_{1,2}+\hat{C}_{2,1}-\hat{C}_{2,2},
\end{eqnarray}
where $\hat{C}_{a,b} =\hat{A}_a\otimes\hat{B}_b$ is the correlation
operator. We call $ {\cal \hat{W}}$ an entanglement witness
operator. Note that the operator in Eq.~(\ref{eq:BOp}) can also be
regarded as a Bell operator ${\cal \hat{B}}$ which distinguishes
non-local properties from local realism. The bound expectation value
of the operator in Eq.~(\ref{eq:BOp}) is determined according to
whether it is regarded as $ {\cal \hat{W}}$ or ${\cal \hat{B}}$. In
other words, the entanglement criterion given by the operator
(\ref{eq:BOp}) is different with the non-locality criterion as we
will show below.

Let us firstly obtain the bound expectation value of the operator
(\ref{eq:BOp}) as an entanglement witness by which one can
discriminate entangled states and separable states. For a separable
state $\hat{\rho}_{\mathrm{sep}}= \sum_i p_i \hat{\rho}_i^{A}\otimes
\hat{\rho}_i^{B}$ where $p_i \geq 0$ and $\sum_i p_i =1$, the
expectation value of the correlation operator measured with
efficiency $\eta$ is given by
\begin{eqnarray}
 \label{eq:expCorr}
\nonumber
  \langle \hat{C}_{a,b} \rangle^{\mathrm{sep}}_{\eta}&=&\sum_i p_i
  \sum^{\infty}_{n,m}(1-2\eta)^{n+m}\\
\nonumber
  &&~~~~~\times \bra{\alpha, n}\hat{\rho}_i^{A}\ket{\alpha, n}\bra{\beta, m}\hat{\rho}_i^{B}\ket{\beta,
  m}\\
  &=&\sum_i p_i \langle \hat{A}_a \rangle^i_{\eta}\langle \hat{B}_b
  \rangle^i_{\eta}.
\end{eqnarray}
Since expectation values of all local observables with efficiency
$\eta$ are bounded as $|\langle \hat{A}_a \rangle^i_{\eta}|,|\langle
\hat{B}_b \rangle^i_{\eta}|\leq 1$ for $a,b=1,2$, we can obtain the
statistical maximal bound of the entanglement witness operator
(\ref{eq:BOp}) with respect to the separable states:
\begin{eqnarray}
 \label{eq:EWb}
\nonumber
  |\langle {\cal \hat{W}} \rangle^{\mathrm{sep}}_{\eta}| &=&
  \biggl|\sum_i p_i (\langle \hat{A}_1 \rangle^i_{\eta}\langle \hat{B}_1
  \rangle^i_{\eta}+\langle \hat{A}_1 \rangle^i_{\eta}\langle
  \hat{B}_2
  \rangle^i_{\eta}\\
\nonumber
  &&~~+\langle \hat{A}_2 \rangle^i_{\eta}\langle
  \hat{B}_1
  \rangle^i_{\eta}-\langle \hat{A}_2 \rangle^i_{\eta}\langle
  \hat{B}_2 \rangle^i_{\eta})\biggr| \\
  &&\leq 2\sum_i p_i =2\equiv {\cal W}^{\mathrm{sep}}_{\mathrm{max}}.
\end{eqnarray}
Therefore, if $|\langle {\cal \hat{W}}
\rangle^{\mathrm{\psi}}_{\eta}| > {\cal
W}^{\mathrm{sep}}_{\mathrm{max}} =2$ for a quantum state $\psi$, we
can conclude that the quantum state $\psi$ is entangled.

Let us then consider the operator (\ref{eq:BOp}) as a Bell operator.
Note that the local-realistic (LR) bound of a Bell operator is given
as the extremal expectation value of the Bell operator, which is
associated with a deterministic configuration of all possible
measurement outcomes. If $1/2 < \eta \leq 1$, the maximal modulus
outcome of (\ref{eq:ObOp}) is $|1-2/\eta|$ when the outcome of
parity operator $\hat{\Pi}(\alpha)$ is measured as $-1$. Thus the
expectation value of (\ref{eq:BOp}) is bounded by local realism as
$|\langle {\cal \hat{B}}\rangle_{\eta}| \leq {\cal
B}^{\mathrm{LR}}_{\mathrm{max}} =2(1-2/\eta)^2$. Likewise for $\eta
\leq 1/2$, we can obtain ${\cal B}^{\mathrm{LR}}_{\mathrm{max}}
=18$. Note that ${\cal B}^{\mathrm{LR}}_{\mathrm{max}} \geq {\cal
W}^{\mathrm{sep}}_{\mathrm{max}}$ for all $\eta$, and ${\cal
B}^{\mathrm{LR}}_{\mathrm{max}} = {\cal
W}^{\mathrm{sep}}_{\mathrm{max}}$ in the case of unit efficiency
($\eta=1$). It shows that some entanglement can exist without
violating local realism, and thus the Bell operator can be regarded
as a non-optimal entanglement witness as pointed out already in
\cite{Hyllus-PRA-2005}. For the purpose of this paper we will focus
on the role of an entanglement witness in the following parts.

From Eq.~(\ref{eq:BOp}) and Eq.~(\ref{eq:EWb}), we can finally
obtain an entanglement witness in the form of an inequality obeyed
by any separable state:
\begin{eqnarray}
\label{eq:BFinSW} \nonumber
  |\langle {\cal \hat{W}} \rangle_{\eta>\frac{1}{2}} |&=&\biggl|
  \frac{\pi^2}{4\eta^2}[W^{\eta}_{1,1}+W^{\eta}_{1,2}+W^{\eta}_{2,1}
  -W^{\eta}_{2,2}]\\
\nonumber
  &+&\frac{\pi(\eta-1)}{\eta^2}[W^{\eta}_{a=1}
  +W^{\eta}_{b=1}]
  +2(1-\frac{1}{\eta})^2 \biggr| ~\leq 2,\\
\nonumber
  |\langle {\cal \hat{W}} \rangle_{\eta\leq\frac{1}{2}} |&=&|\pi^2[W^{\eta}_{1,1}
  +W^{\eta}_{1,2}+W^{\eta}_{2,1}-W^{\eta}_{2,2}]\\
  &&~~~~-2\pi[W^{\eta}_{a=1}+W^{\eta}_{b=1}]+2|~\leq
  2,
\end{eqnarray}
where $W^{\eta}_{a,b}$ is the two-mode Wigner function measured with
efficiency $\eta$ (here we replace the notation $\alpha_a$ and
$\beta_b$ in the conventional representation of two-mode Wigner
function $W^{\eta}(\alpha_a,\beta_b)$ with the notation $a,b$ for
simplicity), and $W^{\eta}_{a(b)=1}$ is its marginal single-mode
distribution. We note that the entanglement witness in
Eq.~(\ref{eq:BFinSW}) can be also derived from the Bell inequality
formulated using $s$-parameterized quasi-probability functions
\cite{SWLEE09} when regarding effects of detector inefficiency as
changes to $s$. Any violation of Eq.~(\ref{eq:BFinSW}) guarantees
that the measured quantum state is entangled. Remarkably, our scheme
allows one to detect entanglement without correcting measurement
errors. Note that in the case of unit efficiency ($\eta=1$) the
inequality in Eq.~(\ref{eq:BFinSW}) becomes equivalent to the
BW-inequality \cite{Banaszek-PRA-1998}. It is also notable that any
violation of this inequality for $\eta < 1$ ensures the violation of
the BW-inequality in the case of a unit efficiency ($\eta =1$).
Therefore the proposed entanglement witness in Eq.~(\ref{eq:BFinSW})
can be used effectively for detecting entanglement instead of the
BW-inequality in the presence of measurement noise.

%

\section{Testing single photon entangled states}

\begin{figure}
\begin{center}
\includegraphics[width=0.45\textwidth]{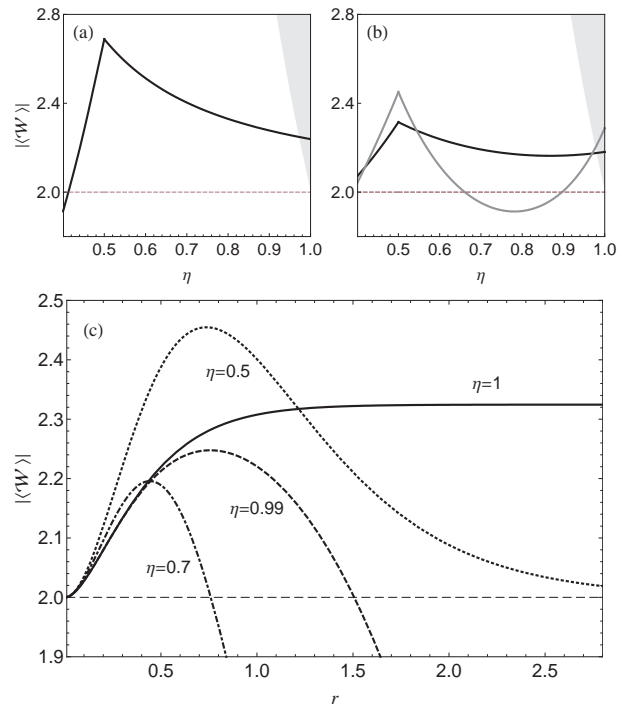}
\caption{Maximum expectation value of the entanglement witness
operator in Eq.~(\ref{eq:BOp}) for an input of (a) a single photon
entangled state and (b) a two-mode squeezed state with $r=0.4$
(black) and $r=0.8$ (grey). Entanglement exists if the expectation
value exceeds the dashed line ${\cal
W}^{\mathrm{sep}}_{\mathrm{max}} =2$. Note that the shaded region
which exceeds ${\cal B}^{\mathrm{LR}}_{\mathrm{max}} =2(1-2/\eta)^2$
(for $1/2 < \eta \leq 1$) is the criterion of non-locality. (c)
Witnessing entanglement with varying squeezing rate $r$ of a
two-mode squeezed states for detector efficiencies $\eta=1$ (solid
line), $\eta=0.99$ (dashed line), $\eta=0.7$ (dotdashed line) and
$\eta=0.5$ (dotted line).}\label{fig6-1}
\end{center}
\end{figure}

Let us now apply the entanglement witness in Eq.~(\ref{eq:BFinSW})
for detecting entangled photons. We here firstly consider the single
photon entangled state $\ket{\Psi}=(\ket{0,1}+\ket{1,0})/\sqrt{2}$
where $\ket{0,1} (\ket{1,0})$ is the state with zero (one) photons
in the mode of Alice and one (zero) photon in the mode of Bob
\cite{singlephotone}. This state can be created by a single photon
incident on a 50:50 beam splitter. Its two-mode Wigner function
measured with efficiency $\eta$ is
\begin{eqnarray}
\nonumber
W^{\eta}_{a,b}=\frac{4}{\pi^2}(1-2\eta+2\eta^2|\alpha_a+\beta_b|^2)\\
~~~~~~~~~~~~~\times\exp[-2\eta(|\alpha_a|^2+|\beta_b|^2)]
\end{eqnarray}
and its marginal single-mode distribution is
\begin{eqnarray}
W^{\eta}_{a}=\frac{1}{\pi}(2-2\eta+4\eta^2|\alpha_a|^2)
\exp[-2\eta|\alpha_a|^2].
\end{eqnarray}
The expectation values of operator (\ref{eq:BOp}) with properly
chosen $\alpha_a$ and $\beta_b$ are plotted in Fig.~\ref{fig6-1}(a)
against the overall efficiency $\eta$. It is remarkable that
entanglement can be detected even with detection efficiency $\eta$
as low as 40\%.

\section{Testing two-mode squeezed states}

Let us consider the entanglement witness in continuous variable
systems e.g. two-mode squeezed states (TMSSs). This state can be
generated by non-degenerate optical parametric amplifiers
\cite{Reid-PRL-1988}, and be written as
$\ket{\mathrm{TMSS}}=\mathrm{sech}~r\sum_{n=0}^{\infty}\tanh^{n}{r}\ket{n,n}$
where $r>0$ is the squeezing parameter. The measured Wigner function
with efficiency $\eta$ for a two-mode squeezed state is given by
\begin{eqnarray}
 \label{eq:QdfTMSS}
 \nonumber
W^{\eta}_{a,b}&=&\frac{4}{\pi^2\eta^2R(\eta)}\exp\biggl(-\frac{2}{R(\eta)}
  \{S(\eta)(|\alpha_a|^2+|\beta_b|^2)\\
  &&~~~~~~~~~~-\sinh{2r}(\alpha_a\beta_b+\alpha_a^*\beta_b^*)\}\biggr),
\end{eqnarray}
and its marginal single-mode Winger function is
\begin{eqnarray}
 \label{eq:sQdfTMSS}
  W^{\eta}_{a}=\frac{2}{\pi \eta S(\eta)}\exp\biggl(-\frac{2|\alpha_a|^2}{S(\eta)}\biggr),
\end{eqnarray}
where $R(\eta)=2(1-1/\eta)(1-\cosh{2r})+1/\eta^2$ and
$S(\eta)=\cosh{2r}-1+1/\eta$. The expectation values of the
entanglement witness operator (\ref{eq:BOp}) for two-mode squeezed
states are shown in Fig.~\ref{fig6-1}(b) with different squeezing
rates $r$. It shows that our scheme allows one to detect some
continuous variable entanglement with detector efficiency of about
40 \%. As shown in Fig.~\ref{fig6-1}(c), violations of the
inequality show different tendencies depending on efficiency $\eta$
with increasing the squeezing parameter $r$. In the case of low
squeezing rates the violation is maximized when $\eta=0.5$, while
for larger squeezing rates about $r \geq 1.2$ the violation is
maximized when $\eta=1$. This is because the dominant degree of
freedom of entanglement detected by observable in
Eq.~(\ref{eq:ObOp}) changes with decreasing the efficiency $\eta$.
Note that in the case $\eta=0.5$ the dominant contribution to the
entanglement arises from quantum correlations between the vacuum and
the photon being present, while for $\eta=1$ it comes from
higher-order correlations of photon number states.

\section{Testing with a priori estimated efficiency}

\begin{figure}
\begin{center}
\includegraphics[width=0.47\textwidth]{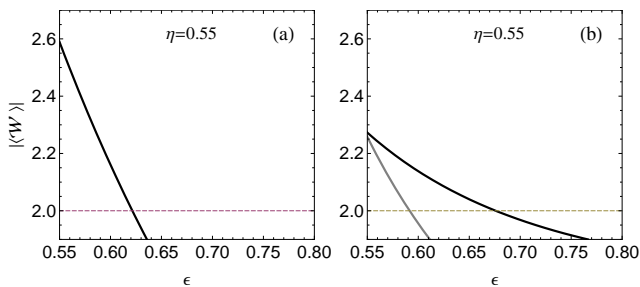}
\caption{Witnessing entanglement with a real efficiency $\eta=0.55$
as varying the estimated efficiency $\varepsilon$ for an input of
(a) a single photon entangled state and (b) a two-mode squeezed
state with $r=0.4$ (black) and $r=0.8$ (grey).}\label{fig6-2}
\end{center}
\end{figure}

So far it has been assumed that the detector efficiency is known
precisely prior to the tests both in our scheme presented above and
in other proposals proposed previously
\cite{Kiss-PRA-1995,Banaszekb-PRA-1998,Lvovsky-RMP-2009}. This can
be realized e.g. by a full characterization of detectors when doing
a quantum tomography on the detectors which has been experimentally
achieved \cite{Lundeen-NatureP-2009}. However, in most cases {\em a
priori} estimates of the detector efficiency ($\equiv \varepsilon$)
may not be perfect and thus can be different from the real
efficiency $\eta$ that affects measured data. Let us assume that we
can discriminate perfectly only whether the real efficiency
$\eta>1/2$ or $\eta\leq1/2$. If $\eta\leq1/2$, we can see that the
entanglement witness in Eq.~(\ref{eq:BFinSW}) is formulated only by
experimentally measured Wigner functions. Thus, in this case our EW
can be tested without knowing the real efficiency. On the other
hand, for the case $\eta>1/2$, the efficiency variable $\eta$ is
explicitly included in the entanglement witness (\ref{eq:BFinSW})
and should be replaced with the estimated efficiency $\varepsilon$
as
\begin{eqnarray}
\label{eq:nBFinSW}
  |\langle {\cal \hat{W}} \rangle_{\eta>\frac{1}{2}} |&=&\biggl|
  \frac{\pi^2}{4\varepsilon^2}[W^{\eta}_{1,1}+W^{\eta}_{1,2}+W^{\eta}_{2,1}
  -W^{\eta}_{2,2}]\\
\nonumber
  &+&\frac{\pi(\varepsilon-1)}{\varepsilon^2}[W^{\eta}_{a=1}
  +W^{\eta}_{b=1}]
  +2(1-\frac{1}{\varepsilon})^2 \biggr| ~\leq 2.
\end{eqnarray}
Note that Eq.~(\ref{eq:nBFinSW}) is valid subject to the condition
\begin{eqnarray}
\nonumber \eta (\mathrm{real~efficiency}) \leq \varepsilon
(\mathrm{estimated~ efficiency}),
\end{eqnarray}
since otherwise the right-hand side of inequality in
Eq.~(\ref{eq:nBFinSW}) is not valid i.e. the expectation values of
separable states are not bounded by ${\cal
W}^{\mathrm{sep}}_{\mathrm{max}} =2$. In this case, one can also
detect entanglement even with non-perfect estimates of the
efficiency as shown in Fig.~\ref{fig6-2}. For example, even when one
estimates the efficiency as $\varepsilon=0.65$ for the detector with
real efficiency $\eta=0.55$, one can still detect entanglement of
the two-mode squeezed state with $r=0.4$ using our scheme.

\section{Conclusions}

We have formulated an entanglement witness that allows one to detect
entanglement even with significantly imperfect detectors. The
proposed entanglement witness in (\ref{eq:BFinSW}) can be used to
test arbitrary quantum states represented in phase space using the
Wigner function. It can be implemented by both homodyne
\cite{Vogel-PRA-1989,Leonhardt-PRA-1993} and number counting
tomography methods \cite{Banaszek-PRL-1996} without additional steps
for correcting measurement errors. Moreover, since the required
minimal efficiency for our scheme is as low as 40\%, it may be
realizable using current detection technologies. In addition, our
entanglement witness can be used without knowing the detection
efficiency precisely prior to the test, which may allow more
realistic implementation. We note that our approach is applicable to
e.g. cavity QED or ion trap systems with the help of the direct
measurement scheme of Wigner function in such systems
\cite{Lutterbach-PRL-1997}. It will be also valuable to apply our
scheme to quantum cryptography in which witnessing entanglement is a
primal step for secure quantum key distribution \cite{Curty04}. We
expect that our scheme enhances the possibility of witnessing
entanglement in complex physical systems using current
photo-detection technologies.

\acknowledgments We thank E. Knill, Y. Zhang and S. Thwaite for
valuable comments. This work was supported by the UK EPSRC through
projects QIPIRC (GR/S82176/01), EuroQUAM (EP/E041612/1), the World
Class University (WCU) program and the KOSEF grant funded by the
Korea government(MEST) (R11-2008-095-01000-0).


\begin{thebibliography}{99}

\bibitem{Amico-RMP-2008} L. Amico, R. Fazio, A. Osterloh, and V.
Vedral, Rev. Mod. Phys. {\bf80}, 517 (2008).
\bibitem{Vedral-Nature-2008} V. Vedral, Nature (London). {\bf453}, 1004 (2008).
\bibitem{Nielsen-Cam-2000} M. A. Nielsen and I. L. Chuang, {\em Quantum
Computation and Quantum Information}, (Cambridge University Press,
cambridge, U.K., 2000).
\bibitem{Horodecki-RMP-2009} R. Horodecki, P. Horodecki, M. Horodecki and K.
Horodecki, Rev. Mod. Phys. {\bf81}, 865 (2009).
\bibitem{Braunstein-RMP-2005} S. L. Braunstein, and P. van Loock, Rev. Mod.
Phys. {\bf77}, 513 (2005).
\bibitem{Vogel-PRA-1989} K. Vogel, and H. Risken, \pra {\bf 40}, 2847
(1989); D. T. Smithey {\em et al.}, \prl {\bf70}, 1244 (1993).
\bibitem{Banaszek-PRL-1996} K. Banaszek, and K. W\'{o}dkiewicz, \prl {\bf76}, 4344
(1996); S. Wallentowitz, and W. Vogel, \pra {\bf53}, 4528 (1996).
\bibitem{Lvovsky-RMP-2009} A. I. Lvovsky, and M. G. Raymer, Rev. Mod. Phys.
{\bf81}, 299 (2009).
\bibitem{Bell-Physics-1964} J. S. Bell, Physics {\bf 1}, 195 (1964).
\bibitem{Banaszek-PRA-1998} K. Banaszek, and K. W\'{o}dkiewicz,  \pra {\bf58}, 4345 (1998);
\prl {\bf82}, 2009 (1999); Acta Phys. Slovaca {\bf49}, 491 (1999).
\bibitem{Kiss-PRA-1995} T. Kiss, U. Herzog, and U. Leonhardt, \pra {\bf52},
2433 (1995).
\bibitem{Banaszekb-PRA-1998} K. Banaszek, \pra {\bf57}, 5013 (1998); A. I.
Lvovsky, J. Opt. B: Quantum Semiclass. Opt. {\bf6}, S556 (2004).
\bibitem{Leonhardt-1997} U. Leonhardt, {\em Measuring the Quantum State of
Light}, (Cambridge University Press, Cambridge, 1997).
\bibitem{Cahill-PR-1969} K. E. Cahill, and R. J. Glauber, Phys. Rev. {\bf 177},
1857 (1969); Phys. Rev. {\bf 177}, 1882 (1969).
\bibitem{Leonhardt-PRA-1993} U. Leonhardt, and H. Paul, \pra {\bf
48}, 4598 (1993).
\bibitem{Hyllus-PRA-2005} P. Hyllus, O. G\"{u}hne, D. Bru{\ss}, and M. Lewenstein, \pra
{\bf72}, 012321 (2005).
\bibitem{SWLEE09} S.-W. Lee, H. Jeong, and D. Jaksch, \pra {\bf80} 022104 (2009).
\bibitem{singlephotone} S. M. Tan, D. F. Walls, and M. J. Collett,
\prl {\bf66}, 252 (1991); L. Hardy, \prl {\bf73}, 2279 (1994).
\bibitem{Reid-PRL-1988} M. D. Reid, and P. D. Drummond, \prl {\bf60}, 2731
(1988).
\bibitem{Lundeen-NatureP-2009} J. S. Lundeen {\em et al.}, Nature Physics {\bf 5}, 27
(2009).
\bibitem{Lutterbach-PRL-1997} L. G. Lutterbach, and L. Davidovich, \prl
{\bf78}, 2547 (1997).
\bibitem{Curty04} M. Curty, M. Lewenstein, and N. L\"{u}tkenhaus, \prl {\bf92}, 217903 (2004).










\end{thebibliography}
\end{document}